\documentclass[]{camera}
\usepackage{graphicx}  
\usepackage{epsfig}

\begin{document}

%
\title{Heavy ion Physics with ALICE}

%
\author{Mercedes L\'opez Noriega for the ALICE Collaboration}

%
\organization{Institut de Physique Nucl\'eaire d'Orsay}

\maketitle

\begin{abstract}
ALICE will study the physics of the strongly interacting matter
produced in nucleus-nucleus collisions at the LHC where the
formation of the Quark Gluon Plasma is expected. The experimental
setup, the capabilities of the detector, and a few selected
heavy-ion topics will be presented and discussed.
\end{abstract}

%
Quantum Chromodynamics (QCD) predicts a transition from a stable
state of matter formed by hadrons to a plasma of deconfined quarks
and gluons at sufficiently high energy density where the average
distance between particles becomes so small that confinement
disappears. Ultrarelativistic heavy ion collisions are the way of
reaching energy densities about the critical one to create this
Quark Gluon Plasma (QGP) in the laboratory. This should allow us to
better understand the strong interaction by studying the properties
of the phase transition and the hadron formation.

This kind of collisions will soon take place at the Large Hadron
Collider (LHC) located at CERN. Heavy ion beams will collide with a
maximum center-of-mass energy of $5.5$ ATeV in the case of PbPb
collisions. This implies an increase on energy of about a factor of
30 with respect to the highest energy collisions that have taken
place until now. As a consequence the energy density reached in
these collisions is expected to be between 3 and 10 times larger.

\section{ALICE}
\label{alice}

ALICE, A Large Ion Collider Experiment, is the only LHC experiment
dedicated to the study of heavy ion collisions~\cite{PPRI}. It
consists of different parts.

The central part is located around the collision point and inside a
solenoidal magnet that will provide a magnetic field up to $0.5$~T.
Several detectors will assure the reconstruction of particles in the
rapidity range $|y|<0.9$ with full azimuthal coverage. Those
detectors are, as seen by a particle traveling out from the
interaction point:
\begin{itemize}
    \item ITS, Inner Tracking System, with an inner radius
    of $4$~cm and an outer radius of $40$~cm, it is formed
    by three subsystems of two layers each: a
    silicon pixel, a silicon drift, and a silicon
    strip detector. It will allow the $3$-D reconstruction of the
    collision primary vertex, secondary vertex finding, and
    particle identification via $dE/dx$.
    \item TPC, Time Projection Chamber, which is the main tracking
    detector for charged particles in the central barrel of ALICE.
    It has an inner radius of $0.9$~m, an outer radius of $2.5$~m,
    and a length of $5.1$~m. It is optimized for large track
    densities (up to $dN/dy$~= 8000) and it will allow track finding,
    momentum measurements, and charged particle identification via
    $dE/dx$.
    \item TRD, Transition Radiation Detector, consists of 18
    longitudinal supermodules (6 ready for the first run, limiting
    the azimuthal acceptance), 6 radial layers and 5 stacks
    along the beam axis. Its prime function is the identification of
    electrons; it will also provide fast triggering.
    \item TOF, Time Of Flight, consists of 18 longitudinal
    modules and 5 modules along the beam axis. It will allow
    charged particle identification.
\end{itemize}
Also in the central part there are some detectors with smaller
coverage. The HMPID, High Momentum Particle IDentification, which
provides particle identification for particles in the $1$ to
$6$~GeV/c momentum range. The PhoS, Photon Spectrometer, and the
EMCal, ElectroMagnetic Calorimeter, for photon and neutral particle
identification.

In the pseudorapidity region $-4.0<\eta<-2.5$ there is a Muon
Spectrometer designed to detect muons at backward rapidity with a
mass resolution of $1\%$ at $10$~GeV/c$^2$. It consists of a large
front absorber located very close to the interaction point to
minimize contributions from hadrons and photons; 5 tracking stations
of two detectors planes each to reconstruct the muons; 2 trigger
stations behind a muon filter that provides fast trigger
capabilities; and a dipole magnet with a field integral of $3$~Tm.

There is also a set of forward detectors close to the LHC beam. The
T0 for event triggering and tagging; the FMD that provides
multiplicity information over $-3.4<\eta<-1.7$ and $1.7<\eta<5.0$;
ZDCs for spectator nucleons and protons; the V0 that provides
triggering and luminosity information; and the PMD to provide
information on photon production at forward rapidities.

These detectors, and combining several identification techniques,
will allow the identification of particles over a wide rapidity
range up to large momentum values.

\section{Heavy-ion physics with ALICE}
\label{hip}

When two heavy-ion nuclei collide, there is a pre-equilibrium state
in which each nucleon scatters several times and partons are
liberated. These quarks and gluons thermalize by re-scattering
resulting in a thermalized QGP. The system then expands collectively
and cools down to temperatures around the critical temperature when
hadrons are formed. The hadrons interact inelastically until the
system reaches what is known as the chemical freeze-out and this
kind of interactions stops. The system keeps expanding and
eventually it is diluted enough that the interactions between
hadrons stop, the system undergoes a thermal freeze-out, and hadrons
fly off to our detectors. Several observables allow us to study the
created medium and the different stages of the evolution of the
system. In this section I will describe some of these observables
and show the expected performance of ALICE.

\subsection{Soft physics: particle ratios and elliptic flow} \label{soft}

We call ``soft physics" to the observables dominated by low $p_T$
particles, \emph{i.e.} the $p_T$ region below $3$ or $4$~GeV/c.

From the chemical composition of the system and its comparison to
equilibrium and non-equilibrium statistical model predictions we
will be able to extract the freeze-out parameters (chemical
freeze-out temperature and baryon-chemical potential) and the degree
of equilibration of the system~\cite{PBraun01,JRafel06}. In
particular, non-identical particle ratios with strange baryon
involved are very sensitive to the equilibrium hypothesis. The
distinction between equilibrium and non-equilibrium scenarios should
be possible in ALICE.

The system created in a non-central collision has an initial
azimuthal anisotropy in the coordinate space. The multiple
interactions among the constituents will generate a pressure
gradient that will transform this initial coordinate space
anisotropy into a momentum space anisotropy. The magnitude of $v_2$,
the second Fourier coefficient of the azimuthal anisotropy of
particle production, and its $p_T$ dependence allow for the
extraction of the thermal freeze-out temperature and transverse flow
velocity. The question whether $v_2$ has reached the hydrodynamical
limit at RHIC, reflecting a perfect fluid behavior, is still highly
debated. According to several predictions one should be closer to
ideal hydrodynamics at the LHC while a significant increase on $v_2$
is still possible. In ALICE, different methods will be applied to
extract $v_2$.

\subsection{Hard probes} \label{hard}

The domination of hard processes, those coming from initial hard
scattering of nucleon constituents, in the early stage of the
collision will have as a consequence that very hard probes will be
copiously produced and therefore event-by-event jet reconstruction
and jet-quenching studies will be possible, and heavy flavors like
beauty will become accessible.

\subsubsection{High-$p_{\textrm{T}}$ physics} \label{highpt}

The large transverse momentum partons generated in the initial hard
scattering of nucleon constituents will fragment to create a high
energy cluster of particles, a jet. These partons will also travel
through what is predicted to be a dense colored medium and there
they are expected to loose energy via medium induced gluon
radiation~\cite{XWang94,XWang95}. The magnitude of this energy loss
is predicted to depend strongly on the gluon density of the medium.
Therefore, measurements on how quenching changes the structure of
the jet and its fragmentation function will reveal information aboutthe QCD medium created in these collisions.

\begin{figure}
\begin{center}
\resizebox{0.70\textwidth}{!}{
\includegraphics{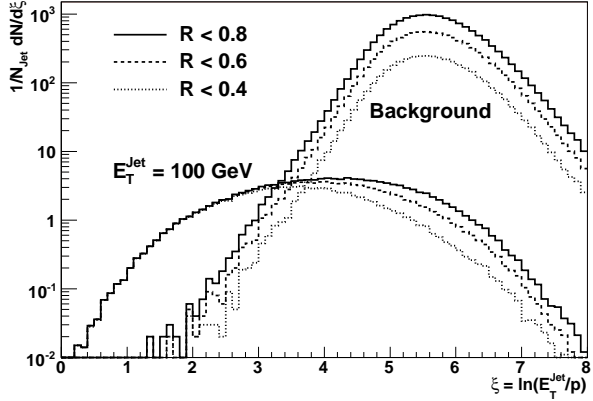}
}\caption{The hump-backed plateau for 100 GeV jets compared to the
contribution from the background of the underlying event for
different cone sizes $R$.} \label{hbp}
\end{center}
\end{figure}

A convenient way of studying how the medium modifies the jet
structure is through the fragmentation function, represented in
Fig.~\ref{hbp} by the distribution of
$\xi=\ln(E_{\textrm{T}}^{\textrm{jet}}/p)$. The particular shape of
the distribution is called the hump-backed plateau~\cite{NBorghi05}.
Medium induced energy loss distorts the shape of the plateau in a
characteristic way. Figure \ref{hbp} shows the hump-backed plateau
for $100$~GeV jets compared to the contribution from the background
underlying event for different cone sizes $R$ (where the jet energy
is contained) as seen in simulations in ALICE. The region $\xi<4$
corresponds to particles with $p_T$ larger than $1.8$ and therefore
the leading particle remanent can be observed with $S/B>0.1$.
Particles from medium-induced gluon radiation are expected to show
up mostly in the region $4<\xi<6$ where the $S/B$ is of the order of
$10^{-2}$.

One very attractive method to obtain an unbiased measurement of the
original parton energy is to tag jets with prompt photons emitted in
the direction opposite to the jet direction. The dominant processes
for such events are Compton scattering and annihilation and they
dominate the photon spectrum at $p_T>10$~GeV/c. This technique helps
to localize the jet. The measured photon energy is equal to that of
the parton before energy loss because photons emerge from the medium
almost unaltered. ALICE will tag jets with photons measured in the
PhoS or the EMCal and will study the parton in-medium-modification
through the fragmentation function.

\subsubsection{Heavy quarks: open heavy flavor and quarkonia} \label{HQ}

Heavy quarks, charm and beauty, will be abundantly produced in
heavy-ion collisions at the LHC, and both the production of open
heavy flavor and quarkonia will probe the strongly interacting
medium created in these collisions. Heavy flavor will also prove the
gluon small Bjorken-x domain where the gluon density is expected to
be close to saturation leading to modifications of the particle
production rate.

The energy loss at the partonic level in heavy-ion collisions is
expected to depend on the color charge (stronger for gluons than for
quarks due to the higher gluon color charge) and on the mass (weaker
for heavy than for light quarks due to the dead cone
effect)~\cite{RBaier97,YDoks01}. One would therefore expect less
high-$p_T$ quenching for heavy flavor particles that originate from
heavy quark jets than for light flavor particles that are created in
both gluons and (mostly) light quark jets. In ALICE we will be able
to separate the production of charm and beauty hadrons and we will
be able to test the energy loss models.

Quarkonia are bound states of heavy quarks. It has been predicted
that if a deconfined medium was created in heavy-ion collisions then
a suppression in the $J/\psi$ (a $c\overline{c}$ bound state)
production would be observed~\cite{TMats86}. This is due to the fact
that in a deconfined medium the attraction between heavy quarks and
antiquarks is reduced due to the color screening induced by lighter
quarks. Results from previous experiments at SPS and RHIC are not
conclusive and puzzling. It is expected that the measurement of the
$J/\psi$ yield in PbPb collisions at $5.5$~TeV will finally
disentangle between the different suppression/regeneration
scenarios. Models that include regeneration predict an enhacement of
$J/\psi$ production at the LHC due to the large number of
$c\overline{c}$ pairs produced. If sequential screening is the right
explanation, $J/\psi$ will finally melt at the LHC. But any of these
scenarios would imply deconfinement of the created system.

\begin{figure}
\begin{center}
\includegraphics{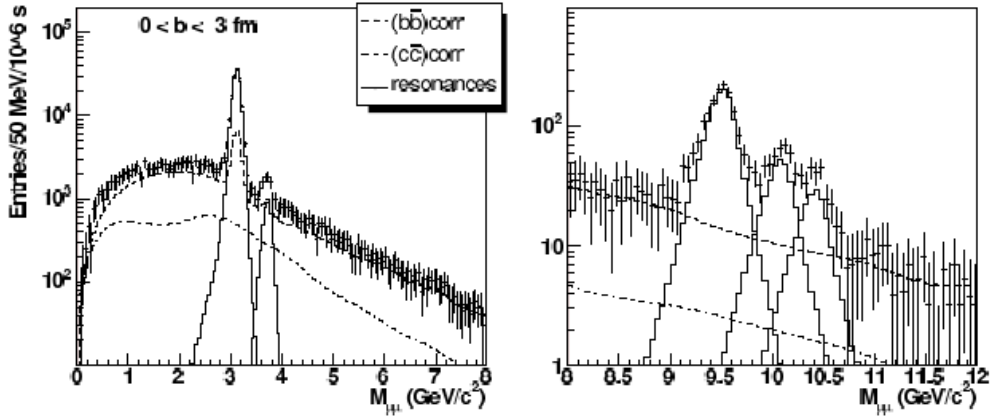}
\caption{Dimuon invariant mass spectra in the $J/\psi$ mass region
(left) and in the $\Upsilon$ mass region (right) after the
subtraction of the uncorrelated background in central PbPb
collisions.}
\label{dimuon} 
\end{center}
\end{figure}

Figure \ref{dimuon} shows the expected ALICE performance for the
measurement of the dimuon spectrum in the muon arm for one month of
PbPb run. ALICE will be able to measure $J/\psi$ with high
statistics in the $p_T$ range $0-20$~GeV/c and therefore we will be
able to measure the predicted effects.

\section{Conclusion}
\label{conclusion}

ALICE is ready for data taking and it is well suited to measure
global event properties on a wide range in PbPb and pp collisions.
The nature of the bulk will be studied via processes like
composition and collective expansion. ALICE will reconstruct jets in
heavy ion collisions to study the properties of the dense created
medium. Charm and beauty production will be studied and the upsilon
family will be accessible for the first time in AA collisions.



%

\end{document}